\documentclass[prd,aps,showpacs,showkeys]{revtex4-1}
\usepackage{bm,amsfonts,latexsym,amsmath,amssymb,amsbsy,graphicx}

\newcommand{\bb}{\begin{eqnarray}}
\newcommand{\ee}{\end{eqnarray}}
\newcommand{\ba}{\begin{align}}
\newcommand{\ea}{\end{align}}

\begin{document}

\title{\bf  Radiative dynamical mass of planar charged fermion in a constant homogeneous magnetic field}
\author{V.R. Khalilov}\email{khalilov@phys.msu.ru}
\affiliation{Faculty of Physics, M.V. Lomonosov Moscow State University, 119991,
Moscow, Russia}

\begin{abstract}
The effective Lagrangian and mass operator are calculated for planar charged  massive and massless fermions in a constant external homogeneous magnetic field in the one-loop approximation of the  2+1 dimensional quantum electrodynamics (QED$_{2+1}$).
We obtain the renormalizable effective Lagrangian and the fermion mass operator for a charged fermion of mass $m$ and then calculate these quantities for the massless case.  The radiative corrections to the mass of charged massless fermion when it occupies  the lowest Landau level are  found for the cases of the pure QED$_{2+1}$ as well as the so-called reduced QED$_{3+1}$ on a 2-brane. The fermion masses were found can be generated dynamically by an external magnetic field in the pure QED$_{2+1}$ if the charged fermion  has small  bare mass $m_0$ and in the reduced QED$_{3+1}$ on a 2-brane even at $m_0=0$. The dynamical mass  seems to be likely to be revealed in monolayer graphene in the presence of constant homogeneous magnetic field (normal to the graphene sample).
\end{abstract}



\maketitle

\section{Introduction}

Quantum systems   of planar charged fermions in external
electromagnetic fields  are interesting in view of possible applications of the corresponding field-theory models to a number of condensed-matter quantum effects such as, for example, the quantum Hall effect \cite{2} and high-temperature superconductivity \cite{3} as well as in connection with problems of
 graphene (see, \cite{6,7,8,9}). In graphene, the electron dynamics
near the Fermi surface  can described by the Dirac equation
in 2+1 dimensions for a zero-mass charged fermion \cite{7} though the case of massive charged fermions is also of interest \cite{13}.
The field-theory models applied for study are  the pure 2+1 dimensional quantum electrodynamics (QED$_{2+1}$) as well as the so-called reduced QED$_{3+1}$ on a 2-brane.
In the latter model,  fermions are confined
to a plane, nevertheless  the electromagnetic interaction
between them is three-dimensional \cite{4,4a}.

The radiative  one-loop shift of an electron energy in the ground state in a constant homogeneous magnetic field in QED$_{2+1}$ was calculated in \cite{TBZh} and the one-loop  electron self-energy in the topologically massive  QED$_{2+1}$ at finite temperature and density was obtained in \cite{ZhEm}.
The effective Lagrangian, the electron mass operator and the density of vacuum electrons (induced by the background field) in an external constant homogeneous magnetic field were derived in the one-loop  QED$_{2+1}$ approximation in \cite{vrkh1}.

Since the effective fine structure constant in graphene is large, the QED$_{2+1}$ effects
can be significant  already in the one-loop approximation.
The polarization operator in graphene in a strong constant homogeneous magnetic field perpendicular to the graphene membrane has been obtained in the one-loop approximation of the  QED$_{2+1}$ in \cite{4a,three,3a,khmam}. The effective potential and vacuum current in graphene in a superposition of a constant homogeneous magnetic field and an Aharonov–Bohm vortex was studied in \cite{StZh}. We also note that the induced vacuum current in the field of a solenoid perpendicular to the graphene sample was investigated in \cite{15},
and the vacuum polarization of massive and massless fermions in an Aharonov--Bohm vortex in the one-loop approximation of the QED$_{2+1}$  was studied in \cite{kh1}.
The one-loop self-energy of a Dirac electron of mass $m$
in a thin medium simulating graphene in the presence of external magnetic field  was investigated in the reduced QED$_{3+1}$ on a 2-brane in \cite{machet}, in which  it was shown that the radiative mass correction  in the lowest Landau level does not vanish at the limit $m\to 0$.

In this work, we calculate the effective Lagrangian and the mass operator of planar charged  fermions in the presence of an external constant homogeneous magnetic field in the one-loop approximation of the QED$2+1$.  We also calculate the radiative corrections to the mass of charged massless fermion when it occupies  the lowest Landau level are  found for the cases of the pure QED$_{2+1}$ as well as the so-called reduced QED$_{3+1}$ on a 2-brane. The fermion masses were found can be generated dynamically by an external magnetic field in the pure QED$_{2+1}$ if the charged fermion  has small  bare mass $m_0$ and in the reduced QED$_{3+1}$ on a 2-brane even at $m_0=0$. The dynamical mass  seems to be likely to be revealed in monolayer graphene in the presence of constant homogeneous magnetic field (normal to the graphene sample) though we also see that an one-loop result is not accurate certainly when the coupling constant is large (see, \cite{machet}).

\section{Eigenfunctions, the Green's function for the Dirac equation in a constant  magnetic field in 2+1 dimensions. Effective Lagrangian}

Since the charged fermions are confined to a plane the exact solutions and the Green's function for the Dirac equation in a constant homogeneous magnetic field can be found in 2+1 dimensions.
The Dirac equation for a fermion in an external electromagnetic field in 2+1 dimensions is  written just as in 3+1 dimensions nevertheless
the Dirac $\gamma^{\mu}$-matrix algebra in 2+1 dimensions is known to be represented
in terms of the two-dimensional Pauli matrices $\sigma_j$
\bb
 \gamma^0= \sigma_3,\quad \gamma^1=i\zeta\sigma_1,\quad \gamma^2=i\sigma_2, \label{1spin}
\ee
where the parameter $\zeta=\pm 1$ can label two types of fermions in accordance with the
signature of the two-dimensional Dirac matrices \cite{27}; it can be applied to characterize two states of the fermion spin (spin "up" and "down") \cite{4c}.

Then, the Dirac equation for a fermion of  mass $m$ and charge $e=-e_0<0$ "minimally" interacting with the background electromagnetic field is written in the covariant form as
\bb
 (\gamma_{\mu}P^{\mu}-m)\Psi(x)=0,\label{dieq}
\ee
where $P^{\mu} = p^{\mu} - eA^{\mu}\equiv (P_0, P_1, P_2)$ is the generalized fermion momentum operator (a three-vector) and $p^{\mu}= (i\partial_t,-i\partial_x,-i\partial_y)\equiv (p_0, p_1, p_2)$. We take the vector potential in the Cartesian coordinates in the Landau gauge   $A_0=0,\quad A_1=0,\quad A_2=Bx$, then the magnetic field is defined as $B=\partial_1A_2-\partial_2A_1\equiv F_{21}$, where $F_{\mu\nu}$ is the electromagnetic field tensor.

The squared Dirac operator is
\bb
(\gamma P)^2= P_0^2-P_1^2-P_2^2 -e\sigma_3 B.
\label{dirop}
\ee
The matrix function $E_p$, introduced by Ritus \cite{virit} for the case of 3+1 dimensions,  satisfies the equation
\bb
(\gamma P)^2E_p= p^2E_p,
\label{epf}
\ee
where the eigenvalue $p^2$ can be any real number and in the magnetic field considered $E_p$ also is an eigenfunction of the operators
\bb
i\partial_t E_p=p_0 E_p,\quad -i\partial_y E_p=p_2 E_p,\quad (P_1^2+P_2^2-e\sigma_3  B)E_p=2|eB|k E_p, \quad k=0, 1 \ldots \quad .
\label{epf1}
\ee
It is obvious that $p_0, p_2$ and $2|eB|n$ label the solutions of Dirac equation (\ref{dieq}) as well as the $E_p$ eigenfunctions of operators (\ref{epf}) and (\ref{epf1}) can also classifies by the eigenvalues $\zeta=\pm 1$ of $\sigma_3$.
The eigenfunctions $E_{p\zeta}(t, {\bf r})$ are given by
\bb
E_{p\zeta}(t, {\bf r})= \frac{1}{2^{3/2}\pi}e^{-ip_0t+ip_2y}U_n(X)w_{\zeta},
\label{epze}
\ee
where the normalized functions $U_n(X)$ are expressed through the Hermite polynomials $H_n(X)$ as
\bb
U_n(X)= \frac{|eB|^{1/4}}{(2^nn!\pi^{1/2})^{1/2}}e^{-X^2/2}H_n(X), X=\sqrt{|eB|}(x-p_2/eB),
\label{ufun}
\ee
$n=k+{\rm sign}(eB)\zeta/2 -1/2,\quad n= 0,1, \ldots ,$ and $w_{\zeta}$ are eigenvectors of $\sigma_3$.

The eigenfunctions of the Dirac equation  Hamiltonian
\bb
H_D=\sigma_1P_2-\sigma_2P_1+\sigma_3m
\label{dihamil}
\ee
are
\bb
 \Psi(t,{\bf r}) =\frac{1}{\sqrt{2E_n}}
\left( \begin{array}{c}
\sqrt{E_n+\zeta m}U_n(X)\\
-{\rm sign}(eB)\sqrt{E_n-\zeta m} U_{n-1}(X)
\end{array}\right)\exp(-iE_nt+ip_2y), \label{three}
\ee
where
\bb
 E_n = \sqrt{m^2+2n|eB|}
\label{Ll}
\ee
is the energy eigenvalues (the Landau levels). All the energy levels except the lowest level ($n=0$) with $\zeta=1$ for $eB>0$ and $\zeta=-1$ for $eB<0$  are doubly degenerate on spin $\zeta=\pm 1$. This means that the eigenvalues of the fermion energy except the lowest level are actually spin-independent in the configuration under investigation. For definiteness, we consider the case where $eB<0$. It should be also emphasized that
$\Psi(t,{\bf r})$ are not eigenfunctions of $\sigma_3$.

The electron propagator in an external constant homogeneous  magnetic field in 3+1 dimensions was found for the first time by  Schwinger \cite{sch}.  In the 2+1 dimensions it was obtained in \cite{vgvmis} in the momentum representation (see, also, \cite{khmam}) and in \cite{vrkh1} in the coordinate (proper-time) representation in the form
\bb
S^c(t,t',{\bf r}, {\bf r}',  B)=-\frac{i^{1/2}}{8\pi^{3/2}}\int\limits_{0}^{\infty}\frac{dz}{s^{1/2}
\sin z}\exp\left[-i(m^2-i\epsilon) -i\frac{(t-t')^2}{4s} +iz\frac{(x-x')^2+(y-y')^2 }{4s\tan z}+ieB\frac{(x-x')(y-y')}{2}\right]\times \nonumber\\
\left[\left(\frac{\gamma^0(t-t')}{2s}+m\right)\exp(i\zeta\sigma_3z)-z\frac{(\gamma^1(x-x')+
\gamma^2(y-y')}{2s\sin z}\right],\phantom{mmmmmmmmm}
\label{maggreen}
\ee
where $z=|eB|s$ and $s$ is the ``proper time".

The  effective action $V(B)=\int d^3x L_{eff}(B)$,  where $L_{eff}(B)$ is
the effective Lagrangian and $x^{\mu}= x^0=t, x^1=x, x^2=y$, is determined  the causal Green's function $S^c$ as follows
\bb
V(B)=i {\rm Tr} \log S^c = i\int d^3 x {\rm tr}(x|\log S^c|x),
\label{eflagr}
\ee
and from Eq. (\ref{maggreen}) we derive
\bb
L_{eff}(B)= \frac{(-i)^{1/2}}{8\pi^{3/2}}\int_0^{\infty}\frac{ds}{s^{3/2}}e^{-im^2s}\left(|eB|\cot|eB|s -\frac1s\right)
\label{lagr0}
\ee
Rotating the integration contour by $-\pi/2$, we have
\bb
L_{eff}(B)= -\frac{|eB|^{3/2}}{8\pi^{3/2}}\int_0^{\infty}\frac{dz}{z^{3/2}}e^{-m^2z/|eB|}\left(\coth z -\frac1z\right)
\label{lagr1}
\ee
and integrating (\ref{lagr1}), one obtains (see, \cite{vrkh1})
\bb
L_{eff}(B)= -\frac{(eB)^2}{24m\pi}\left(1-\frac{(eB)^2}{20m^4}\right),\quad |eB|\ll m^2
\label{weaklagr}
\ee
and
\bb
L_{eff}(B)\approx \frac{(eB)^2}{12m\pi^{3/2}}, \quad |eB|\gg m^2.
\label{limlagr}
\ee
In Eq. (\ref{weaklagr}) the first term can be interpreted as the magnetic energy of the vacuum \cite{nesch}. In the QED$_{3+1}$ the nonrenormalized effective Lagrangian  involves a similar $B^2$ term with a logarithmic factor whose argument depends on the cutoff parameter. This term is then combined  with the $B^2$ term in the initial Lagrangian $L_0(B^2)$, which implies fermion-charge and external magnetic-field renormalizations.

For massless case it is convenient to write the one-loop correction $L_1$ to the Lagrangian density $L_0=-B^2/4\pi$ for the magnetic field as
\bb
L_1 = \frac{|eB|}{2\pi}\sum_{n=1}^{\infty}|E'_n|,
\label{lagr2}
\ee
where the factor $|eB|/2\pi$ takes into account quantum degeneracy of Landau levels per
unit surface area, $E'_n=\sqrt{2|eB|n}$ and the zero modes are to be omitted (see, \cite{kavozu}). A massless fermion does not have a spin degree of freedom in 2+1 dimensions \cite{jacna} but  the Dirac equation  for charged massless fermions in an external magnetic field in 2+1 dimensions keeps the spin parameter. Therefore, all the energy levels except the lowest level $n=0$ are also doubly degenerate. This sum is divergent.
Using for $|E'_n|$ the Fock-Schwinger proper-time representation in the form \cite{hug}
\bb
\sqrt{2|eB|n} = -(\pi)^{-1/2} \int_{0}^{\infty}\frac{ds}{s^{1/2}}\frac{d}{ds} e^{-2|eB|ns}
\label{prop1}
\ee
and applying for (\ref{lagr2}) the zeta-function regularization \cite{Hawk,kavozu},
after some calculations one can bring Eq. (\ref{lagr2}) formally to the form
\bb
L_1 = - \frac{(eB)^{3/2}}{2\sqrt{2}\pi^{3/2}}\int_{0}^{\infty}\frac{x^{-3/2}}{e^x-1}dx \equiv - \frac{(eB)^{3/2}}{2\sqrt{2}\pi^{3/2}}\Gamma(-1/2) \zeta(-1/2),
\label{lagr3}
\ee
where $\Gamma(z)$ and $\zeta(z)$ are respectively the gamma and $\zeta$-functions.
Then using the Riemann identity \cite{BatErd}
\bb
\zeta(z)= \Gamma(1-z)2^z\pi^{z-1}\cos[\pi(1-z)/2]\zeta(1-z),
\label{rel1}
\ee
we obtain the renormalizable effective Lagrangian for charged massless fermions in the final form :
\bb
L_1 = - \frac{(eB)^{3/2}}{4\sqrt{2}\pi^2}\zeta(3/2).
\label{lagr4}
\ee
This formula coincides with the corresponding result obtained in \cite{kavozu}.

\section{Mass operator of a charged fermion in a constant homogeneous magnetic field}

In the coordinate representation, the mass operator of a charged massive fermion in the considered external field in 2+1 dimensions is given by
\bb
M(x,x')= ie^2\gamma^{\mu}S^c(x,x')\gamma^{\nu}S_{\mu\nu}(x-x') - ie^2\gamma^{\mu}S_{\mu\nu}(x-x') {\rm tr}(\gamma^{\nu} S^c(x,x')),
\label{MO1}
\ee
where $S_{\mu\nu}(x-x')$ is the photon propagation function. We note that in the
 QED$_{3+1}$, the second term in the right-hand side of Eq. (\ref{MO1}) is absent in the
constant homogeneous electromagnetic field because the induced vacuum electric current vanishes. In the pure QED$_{2+1}$, the Chern-Simons field dynamically generated in the effective Heisenberg-Euler Lagrangian by the background magnetic field contributes in this second term in the model under consideration. We shall not consider this term.

We shall use the usual photon propagator in 2+1 dimensions (in the Feynman gauge)
\bb
\frac{1}{k^2-i\epsilon} = i \int_{0}^{\infty}du e^{-i(k^2-i\epsilon)},
\label{phusu}
\ee
as well as the "effective" internal photon propagator for
the reduced QED$_{3+1}$ on a 2-brane in which
the photon is allowed to also propagate in the "bulk" \cite{4}
\bb
\frac{1}{\sqrt{k^2-i\epsilon}} = \sqrt{\frac{i}{\pi}} \int_{0}^{\infty}du \frac{e^{-i(k^2-i\epsilon)}}{\sqrt{u}},
\label{phred}
\ee
In the coordinate (proper-time) representation the photon propagator in 2+1 dimensions  reads as
\bb
S_{\mu\nu}(t-t',{\bf r-r}')= g_{\mu\nu}\frac{i^{1/2}}{8\pi^{3/2}}
\int\limits_{0}^{\infty}\frac{du}{u^{3/2}}
\exp\left[-\epsilon u +i\frac{(t-t')^2-(x-x')^2-(y-y')^2 }{4u}\right].
\label{ph21}
\ee
The internal photon propagator for the reduced QED$_{3+1}$ on a 2-brane is given by Eq. (\ref{ph21}) in which $i^{1/2}/8\pi^{3/2}$ and $u^{-3/2}$ are replaced by  $i/8\pi^{1/2}$ and $u^{-2}$, respectively.

We determine the mass operator in the $E_p$ representation as
\bb
M(p,p',B)=\int d^3z \int d^3z' \bar E_p(z)M(z,z',B)E_p'(z'),
\label{mop1}
\ee
where $\bar E_p=\gamma^0E_p^{\dagger}\gamma^0$ and $E_p^{\dagger}$ is the Hermitian conjugate matrix function. The mass operator is diagonal in the $E_p$
representation
\bb
M(p,p',B)=\delta^3(p-p')M(p,B).
\label{mop2}
\ee

We now calculate the mass operator in the special reference frame with using the fermion
propagator (\ref{maggreen}), the photon propagator (\ref{ph21}) and $E_p$-matrix function
(\ref{epze}). It is convenient to introduce variables $t_-=t-t'$, $x_-=x-x'$, $y_-=y-y'$,$t_+=(t+t')/2$, $x_+=(x+x')/2$, $y_+=(y+y')/2$. It is seen that the $t_+$ and $y_+$ integrations give 2$\pi\delta(p_0-p'_0)$ and 2$\pi\delta(p_2-p'_2)$. We then use the
arising $\delta$-functions, integrate with respect to $t_-$ and obtain in the terms independent of $t_-$ the factor
$$
2\sqrt{\frac{\pi su}{i(s+u)}}\exp\left(\frac{ip_0^2 su}{s+u}\right).
$$
The first-order terms in $t_-$ then contain an extra factor $2p_0 su/(s+u)$.

We integrate over $x_+$ with applying formula
\bb
\int_{-\infty}^{\infty}\sqrt{eB}\exp(-iy_-x)U_n(\eta)U_k(\eta') dx = \exp[i(n-k))v]I_{nk}(z),
\label{int5}
\ee
where $U_l(\eta)$ is given by Eq. (\ref{ufun}), $\eta=\sqrt{eB}(x+x_-),\quad \eta'=\sqrt{eB}(x-x_-), \quad v= \arctan(y_-/x_-)$, and $I_{nk}(z)$ is the Laguerre function
in $z=|eB|(x_-^2 + y_-^2)/2$, related to the Laguerre polynomial
$$
L^l_s(z)=e^{z} z^{-l}\frac{d^s}{dz^s}(z^{s+l}e^{-z})
$$
as
\bb
I_{nk}(z)=\frac{1}{\sqrt{n!k!}}e^{-z/2} z^{(n-k)/2} L^{n-k}_k(z).
\label{int4}
\ee
The integration over $x_-$ and $y_-$ is performed in polar coordinates with using the formulas
\bb
\int_{0}^{2\pi} d\phi e^{i(n-k)\phi}  = 2\pi \delta_{nk}, \phi=\arctan(y_-/x_-), \quad
\int_{0}^{\infty} dz e^{-az} L_n^0(z) = \frac{(a-1)^n}{a^{n+1}}, {\rm Re}(a)>0.
\label{frmx-}
\ee
After these integrations the mass operator (\ref{mop2}) at $su/(s+u)\to 0$
has the ultraviolet divergency of the vacuum origin and needs in the renormalization.
The renormalized mass operator in an external electromagnetic field can be represented in the form
\bb
M_{r}(p,B) = M(p,B)-M^0 + M^0-(M^0)_{\gamma p=m} - \left(\frac{\partial M^0}{\partial\gamma p}\right)_{\gamma p=m}(\gamma p-m) \equiv M_c(p,B)+ M^0_{r},
\label{mop3}
\ee
where $M^0=M(p, B=0)$,  the expression $M_c(p,B)=M(p,B)-M^0$ does not contain the divergences and $M^0_{r}$ is the regularized mass operator of fermion in the vacuum, which
vanishes on the mass shell.

As a result of this renormalization, we obtain the renormalized mass operator with the photon propagator (\ref{ph21}) in the form
\bb
M_c(p,B) =-e^2\frac{i^{3/2}}{8\pi^{3/2}}\int_0^{\infty}ds\int_0^{\infty}du\left[ \frac{|eB|}{\Delta (s+u)^{1/2}\sin(|eB|s)}\exp[-ism^2+ip_0^2su/(s+u) -2in\phi]\times \right.\nonumber\\
\left. \times[(3MI                                                                                                                               -\gamma^0P_0)e^{-i\phi}-\gamma^2P_2] -\frac{\exp[-ism^2+ip_0^2su/(s+u)]}{ (s+u)^{3/2}}[3mI-\sigma_3\frac{p_0u}{s+u}]\right],\phantom{mmmm}
\label{mop4}
\ee
where $\Delta=\sqrt{(eBu)^2 +(1+eBu\cot(eBs))^2}, \quad \phi=\arctan[eBu/(1+eBu\cot(eBs))]$, $I$ is the unit two-column matrix, and
\bb
M=i\frac{p_0 u}{s+u} \sin(|eB|s) +m\cos(|eB|s), \quad P_0=\frac{p_0 u}{s+u} \cos(|eB|s) +im\sin(|eB|s),\nonumber\\
 P_2=\sqrt{2|eB|n}\frac{eBu}{\Delta \sin(|eB|s)}. \phantom{mmmmmmmmmmmmmmmmm}
\label{not1}
\ee

Putting $s'=|eB|s,\quad u'=|eB|u$, we obtain
\bb
M_c(p,B) =-e^2\frac{i^{3/2}}{8\pi^{3/2}\sqrt{|eB|}}\int_0^{\infty}ds'\int_0^{\infty}du'
e^{[-is'm^2/|eB|+ip_0^2s'u'/|eB|(s'+u')]} \nonumber\\  \left[\left((3MI-\sigma_3P_0)\exp[-i(2n+1)\phi]-i\sigma_2\frac{\sqrt{2|eB|n}u'\exp[-i2n\phi]}
{\Delta'}\right)\frac{1}{\Delta' \sqrt{s'+u'}}- \right. \nonumber\\
\left. -\left[3mI-\sigma_3\frac{p_0u'}{s'+u'}\right]\frac{1}{(s'+u')^{3/2}}\right],\phantom{mmmmmmmmmmmmmm}
\label{mop5}
\ee
where $\Delta'=\sqrt{u'^2 +2u'\sin s'\cos s' + \sin^2s'},\quad \phi=\arctan[u'/(1+u'\cot(s'))]$.

Now, let us perform a known change of variables (see \cite{ditreu}, p. 44, Eq. (3.12) and, also, \cite{machet})   $s'=sv,\quad u'=s(1-v)$, and introduce a variable $z=isv^2$, which, in fact, is equivalent to the turning of integration path over $s$  on  $\pi/2$ (a Wick rotation). We shall calculate the mass correction for the (on mass-shell)  level $p_0=m,\quad n=0$. As a result of all these transformations, the mass operator can be written as
\bb
M_c(B) =\frac{e^2m}{8\pi^{3/2}}\int_0^{\infty}dz \sqrt{z}e^{-(z/l)}\int_0^1\frac{dv}{v}
\left[\left( \begin{array}{cc} 4-2v+2v\exp[-2z/v]& 0 \\
0& 8-4v+2v\exp[-2z/v]
\end{array}\right)\times \right.\nonumber\\
\left.\times\frac{1}{2z(1-v)+v^2(1-\exp[-2z/v])}-\left( \begin{array}{cc} 2+v & 0 \\
0& 4-v
\end{array}\right)\frac{1}{z}\right],\phantom{mmmmmmmmmmm}
\label{mop6}
\ee
where $l=|eB|/m^2$.
Then, taking into account Eq. (\ref{three}) for the mass correction $\delta m$ of fermion in the lowest Landau level $n=0, \zeta=-1$, we obtain
\bb
\delta m =\frac{e^2m}{8\pi^{3/2}}\int_0^{\infty}dz \sqrt{z}e^{-(z/l)}\int_0^1\frac{dv}{v}
\left[\frac{8-4v+2v\exp[-2z/v]}{2z(1-v)+v^2(1-\exp[-2z/v])}-\frac{4-v}{z}\right].
\label{mop7}
\ee
The mass correction $\delta m_r$ of fermion in the level $n=0, \zeta=-1$ for the reduced QED$_{3+1}$ on a 2-brane is given by
\bb
\delta m_r =\frac{e^2m}{8\pi}\int_0^{\infty}dz e^{-(z/l)}\int_0^1\frac{dv}{\sqrt{1-v}}
\left[\frac{8-4v+2v\exp[-2z/v]}{2z(1-v)+v^2(1-\exp[-2z/v])}-\frac{4-v}{z}\right].
\label{mop8}
\ee

It is useful the integration over $z$ in the limits $[0,\infty)$ to divide into two regions $[0, z_0]$ plus $[z_0,\infty)$ \cite{janc} (see, also, \cite{machet}) with $z_0\sim 1$. Then, when integrating into region $[z_0,\infty)$ the exponentials can be neglected and
integral between the limits $0$ and $z_0$ is small  compared with that in the limits
$[z_0,\infty)$  and it can be neglected.
Having made the integrations over $v$ we obtained
\bb
\delta m =\frac{e^2m}{8\pi^{3/2}}\int_{z_0}^{\infty}dz e^{-(z/l)}
\left(\frac{4}{\sqrt{z}}+\frac{\ln(2z)}{2\sqrt{z}}\right)
\label{mop7a}
\ee
and
\bb
\delta m_r =\frac{e^2m}{8\pi}\int_{z_0}^{\infty}dz e^{-(z/l)}\left(\frac{2\sqrt{2}\pi}{\sqrt{z}}-\frac{20}{3z}\right).
\label{mop8a}
\ee
It should be emphasized that though the first term in integrand of (\ref{mop8a}) has
the form $2\sqrt{2}\pi/\sqrt{z}$ at $z\gg 1$ the exact calculation of corresponding integral over $v$ at  $z=1$ gives $2\sqrt{2}\pi$.

Finally, integrating over $z$ we find the mass corrections in the form
\bb
\delta m =\frac{e^2}{16\pi}\sqrt{|eB|}\left[\ln\left(\frac{|eB|}{m^2}\right)+8\right]
{\rm erfc} \left(\frac{1}{\sqrt{l}}\right)
\label{mop7b}
\ee
and
\bb
\delta m_r =\frac{e^2}{4\pi}\sqrt{|eB|}\left[\sqrt{2}\pi^{3/2}{\rm erfc}\left( \frac{1}{\sqrt{l}}\right)
-\frac{10}{3\sqrt{l}}\Gamma\left(0,\frac{1}{l}\right)\right],
\label{mop8b}
\ee
where (see, for example, \cite{olw})
\bb
{\rm erfc} (z) = \frac{2}{\sqrt{\pi}}\int_{z}^{\infty} e^{-t^2} dt, \quad \Gamma(a,z)=\int_{z}^{\infty} e^{-t} t^{a-1} dt
\label{not4}
\ee
and we put $z_0=1$. In the limit $l\to \infty$
\bb
{\rm erfc}\left(\frac{1}{\sqrt{l}}\right) \approx 1- \frac{2}{\sqrt{\pi l}}, \quad \Gamma\left(0,\frac{1}{l}\right)\approx \ln l - {\cal C}.
\label{not5}
\ee
Here ${\cal C}=0.57721$ is the Euler constant \cite{GR}.

At the limit $m\to 0$, we obtain
\bb
\delta m =\frac{e^2}{16\pi}\sqrt{|eB|}\left[\ln\left(\frac{|eB|}{m_0^2}\right)+8\right], \quad m_0\to 0
\label{mop7c}
\ee
and
\bb
\delta m_r =\frac{e^2}{2\sqrt{2}}\sqrt{\pi|eB|}.
\label{mop8c}
\ee
Eq. (\ref{mop8c}) coincides with Eq.(63) obtained in \cite{machet}.

It should be noted that Eqs. (\ref{mop7c}) and (\ref{mop8c}) describe the main terms of  radiative mass corrections in massive QED$_{2+1}$ and reduced QED$_{3+1}$ on a 2-brane, respectively, in the limits $|eB|\gg m^2$. We see these corrections differ, in principle, from the mass correction in standard massive QED$_{3+1}$ in the limits $|eB|\gg m^2$ (see, \cite{tern,janc}), in which the leading term is determined by
\bb
\delta m_{3+1} =\frac{e^2}{4\pi}m \ln^2\left(\frac{|eB|}{m^2}\right).
\label{mc3+1}
\ee
Eqs. (\ref{mop7c}) and (\ref{mop8c}) show that, in the presence of external magnetic field,  the fermion mass can be generated dynamically in QED$_{2+1}$ at small bare mass $m_0$ and in reduced QED$_{3+1}$ on a 2-brane even at $m_0=0$. Thus, the latter model, in an external constant homogeneous magnetic field, cannot stay massless in the one-loop approximation \cite{machet}.

\section{Resume}

In this work we calculated the effective Lagrangian and the mass operator of planar  charged  fermions in the presence of an external constant homogeneous
magnetic field in the one-loop  approximation of the QED$_{2+1}$. We also calculated
the mass corrections of charged fermion in the lowest Landau level. It is shown that the fermion mass can be generated dynamically by an external constant homogeneous magnetic field in the pure QED$_{2+1}$ if the charged fermion  has small  bare mass $m_0$ and in the reduced QED$_{3+1}$ on a 2-brane even at $m_0=0$.  The dynamical mass  seems to be likely to be revealed in monolayer graphene in the presence of constant homogeneous magnetic field (normal to the graphene sample).

\end{document}